\documentclass[twocolumn,preprintnumbers,aps]{revtex4}
\usepackage{graphicx}
\usepackage{makeidx}

\newcommand{\be}{\begin{eqnarray}}
\newcommand{\ee}{\end{eqnarray}}

\begin{document}

\title{Coincident-Frequency Entangled Photons in a Homogenous Gravitational Field - A Thought Experiment}

\author{Clovis Jacinto de Matos$^{\rm 1}$}

\affiliation{$^{\rm 1}$European Space Agency, 8-10 rue Mario
Nikis, 75015 Paris, France}

\date{\today}

\preprint{}

\begin{abstract}
Assuming that the Principle of energy conservation holds for
coincident-frequency entangled photons propagating in a
homogeneous gravitational field. It is argued that in this
physical context, either Quantum entanglement or the weak
equivalence principle are broken by the photons.
\end{abstract}

\maketitle

{\sc Introduction ---}After reviewing the relativistic effect of
the gravitational redshift for a photon propagating in a
homogeneous gravitational field, a similar experiment is idealized
for the case of two coincident-frequency entangled photons
propagating in a homogeneous gravitational field. The simultaneous
validity of the principle of energy conservation, the weak
equivalence principle, and the entanglement properties are
investigated for the entangled system. Ultimately one concludes
that quantum coherent systems and classical systems cannot
simultaneously comply with the principle of equivalence and
possess entanglement properties, if the principle of energy
conservation is applicable on average to both sets of physical
systems.

\begin{figure}[h!]
\begin{center}
\includegraphics[scale=0.25]{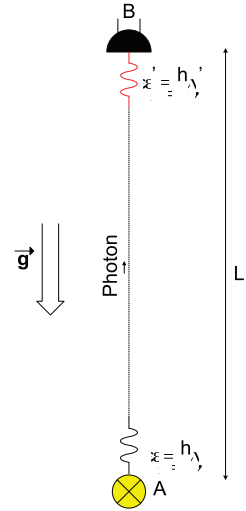}
\caption{\label{fig1} The frequency of a photon propagating
against a homogenous gravitational field $\textbf{g}$ decreases,
in order to comply with the law of energy conservation.}
\end{center}
\end{figure}

{\sc Gravitational Frequency Shift of Light in a Homogeneous
Gravitational Field ---} The law of the gravitational redshift can
be derived directly from the principle of energy conservation
applied to a photon moving against a homogeneous gravitational
field $\vec g$, cf. Fig.\ref{fig1}. A photon source located at
point $A$ where the gravitational potential energy is by
convention set to zero, emits a photon with frequency $\nu$ and
total electromagnetic energy $\epsilon=h\nu$. The photon
propagates against the gravitational field $\vec g$. As it moves
away (along $z$ direction) from the source its gravitational
potential energy, $\epsilon_g=m_{g} g z$, increases; and its
electromagnetic energy, $\epsilon'_1=h\nu'$ must decrease
accordingly in order to maintain the total energy of the photon
equal to its initial value. This should be verified until the
photon reaches the spectrometer located at point $B$ at a distance
$z=L$ above $A$.
\begin{equation}
h\nu=h\nu'+m_{g} g L\label{g1}
\end{equation}
Where $g$ is the module of the gravitational field $\vec g$, and
$m_{g}$ is the photon gravitational mass, which is equal to its
inertial mass $m_{i}$ on the bases of the weak equivalence
principle.
\begin{equation}
m_g=m_i=\frac{h\nu'}{c^2}\label{g2}
\end{equation}
After substituting Equ.(\ref{g2}) in Equ.(\ref{g1}) one can
calculate the relative frequency shift of the photon.
\begin{equation}
\frac{\nu - \nu'}{\nu'}=\frac{gL}{c^2}\label{g3}
\end{equation}
Equ.(\ref{g3}) indicates that the photon's frequency decreases
(redshift) as it moves against the gravitational field $g$. Of
course a blueshift appears for photons path collinear  with
respect to $g$ (from $B$ to $A$).

In the theory of general relativity the photon gravitational
redshift is attributed to a slowing down of clock's frequency with
a reduction of the gravitational field. This is usually derived
from the assumption that the interval $ds^2=c^2dt^2$ connected
with the period of oscillation of an atom $dt$ (considering that
the space coordinates of the atom are fixed), remains unchanged if
the atom is put into a gravitational field with Schwarzschild
metric :
\begin{equation}
ds^2=\Big(1-\frac{2GM}{c^2 R}\Big)c^2dt'^2\label{g4}
\end{equation}

The relativistic gravitational redshift has been experimentally
observed by Pound and Rebka in 1960 in the Earth laboratory, by
measuring the gravitational frequency shift on gamma rays
propagating in a 22.5 meters hight tower \cite{Pound}.

\begin{figure}[h!]
\begin{center}
\includegraphics[scale=0.3]{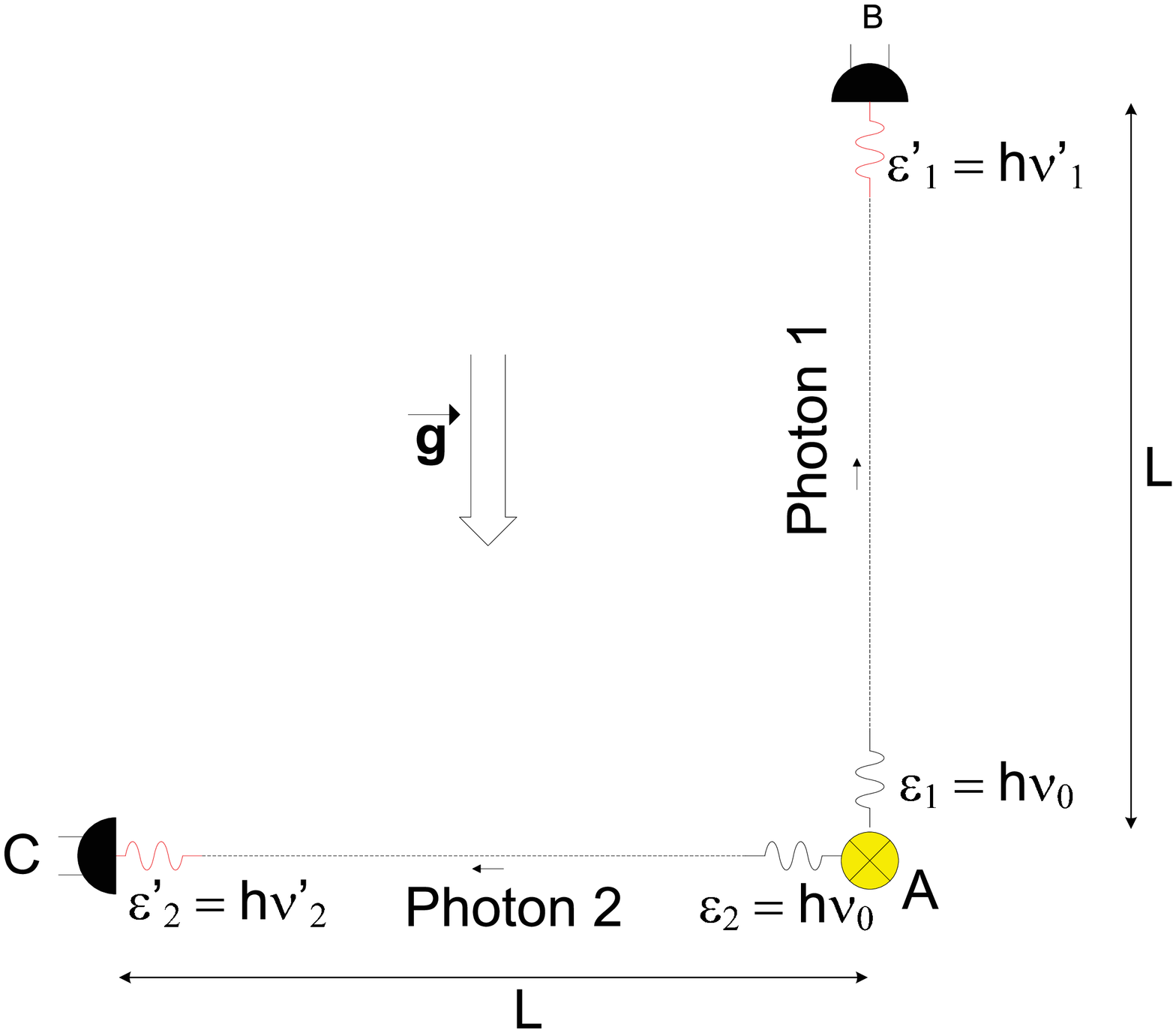}
\caption{\label{fig2} Two coincident-frequency entangled photons
propagating in different gravitational potentials either lose
entanglement, break the weak equivalence principle or violate the
law of energy conservation.}
\end{center}
\end{figure}

{\sc Coincident-Frequency Entangled Photons in a Homogeneous
Gravitational Field ---} Let us modify the traditional
gravitational frequency shift experiment, described schematically
in Fig.\ref{fig1}, by considering two coincident-frequency
entangled photons emitted simultaneously at point $A$ with
frequencies $\nu_0$, instead of simply one single photon. In this
new version of the experiment Photon 1 is propagating along the
vertical direction $z$ against a homogeneous gravitational field
$\vec g$, photon 2 is propagating in the horizontal direction $x$,
orthogonal to $\vec g$, cf. Fig.\ref{fig2}.

Two coincident-frequency entangled photons,
\begin{equation}
|\psi\rangle=\int d\nu \phi(\nu)
|\bar{\nu'_1}+\nu\rangle_1|\bar{\nu'_2}+\nu\rangle_2\label{c1}
\end{equation}
consists of a pair of entangled photons with identical frequencies
$\bar{\nu'_1}=\bar{\nu'_2}$; the two photons are positively
correlated in frequency, and hence anti-correlated in
time\cite{Kuzucu}.

Photon 1 is subject to the relativistic gravitational redshift.
Its frequency decreases according to Equ. (\ref{g3}). Since Photon
1 and 2 have entangled frequencies the frequency of Photon 2
should also decrease as Photon 1 is approaching point B. Although
the gravitational potential energy of photon 2 is kept constant
(equal to zero, since $AC$ defines the level of zero gravitational
potential energy), its electromagnetic energy would be decreasing.
Therefore, if the frequency entanglement between photon 1 and 2 is
preserved while photon 1 propagates against the homogeneous
gravitational field $\vec g$, then photon 2 should violate the
principle of energy conservation.

Imposing that the principle of energy conservation should not be
violated by the entangled photons, then the energy of photon 1 and
2 at point A should remain constant at any time posterior to the
photons emission until their detection at points $B$ and $C$.
\begin{equation}
2\,h\nu_0=h\nu'_1+ h \nu'_2+ m_{g1} g L\label{c2}
\end{equation}
Since it is assumed that the weak equivalence principle is holding
for the photons, one has for photon 1:
\begin{equation}
m_{g1}=m_{i1}=\frac{h\nu'_1}{c^2}\label{c3}
\end{equation}
Substituting in Equ.(\ref{c2}) and solving with respect to
$\nu'_2$ one obtains:
\begin{equation}
\nu'_2=\nu_0\label{c4}
\end{equation}
This shows that the constraint imposed by the quantum entanglement
on the frequencies, $\nu'_1=\nu'_2$, must be relaxed if one
assumes that the principle of energy conservation and the weak
equivalence principle are both holding.

If one now assumes that the principle of energy conservation and
that entanglement must be preserved, then the constraint of having
$\nu'_1=\nu'_2$ in Equ.(\ref{c2}) leads to the following equation
for energy conservation.
\begin{equation}
2\,h\nu_0=2h\nu'_1+ m_{g1} g L\label{c5}
\end{equation}
The adjustable variable in Equ.(\ref{c5}) is now photon 1
gravitational mass. Solving the equation with respect to this
variable.
\begin{equation}
m_{g1}=\frac{2h (\nu_0-\nu'_1)}{gL}\label{c6}
\end{equation}
Photon 1 inertial mass, $m_{i1}$, remains:
\begin{equation}
m_{i1}=\frac{h\nu'_1}{c^2}\label{c7}
\end{equation}
Dividing Equ.(\ref{c6}) by Equ.(\ref{c7}) one estimates any
possible violation of the weak equivalence principle for photon 1
as being proportional to the gravitational redshift:
\begin{equation}
\frac{m_{g1}}{m_{i1}}=2\frac{c^2}{gL}\Big(\frac{\nu_0-\nu'_1}{\nu'_1}\Big)\label{c8}
\end{equation}
If the classical gravitational redshift is observed in this
system, then substituting Equ.(\ref{g3}) in Equ.(\ref{c8}) one
deduces that the gravitational mass of photon 1 is the double of
its inertial mass, between its emission and its reception
\footnote{This result is coherent with the experimental results on
the deflection of light by the Sun, which indicate a gravitational
photon mass equal to the double of its classical inertial value.
Could this mean that every photon in the universe is part of a
correlated photon pair?}.
\begin{equation}
\frac{m_{g1}}{m_{i1}}=2\label{c9}
\end{equation}

{\sc Discussion and Conclusions ---} In the previous section one
has demonstrated that two coincident-frequency entangled photons
propagating in a homogenous gravitational field along
non-collinear optical paths, cannot comply simultaneously with the
principle of energy conservation and the weak equivalence
principle. Thus assuming that the principle of energy conservation
is preserved on average by entangled systems, then either the
entanglement between the photons is lost \cite{Ralph} or the weak
equivalence principle is violated at least for one the photons
according to Equ.(\ref{c8}).

Assuming the universality of the principle of energy conservation,
one concludes that quantum entanglement and the weak equivalence
principle cannot hold simultaneously in a physical system. Thus we
are left with the physical possibilities outlined in table 1,
which divides the physical possible systems in two major sets: On
the one side, \textbf{coherent systems}, like for example:
superconductors, superfluids, Bose Einstein Condensates, Entangled
photons, Entangled quantum bits, which would exhibit macroscopic
and / or  microscopic entanglement, and would violate the weak
equivalence principle. On the other side \textbf{classical
systems}, made out of macroscopic material bodies which do not
possess any form of quantum entanglement between their different
building blocks, but do comply with the weak equivalence
principle.
\begin{center}
\begin{tabular}{|c|c|c|c|}
\hline   & Entanglement & Equivalence Principle \\
\hline Coherent systems & Yes & No \\
\hline Classical systems & No & Yes \\
\hline
\end{tabular}
\end{center}
Table 1: Discriminating between classical and quantum coherent
physical systems using entanglement properties and the weak
equivalence principle.
\bigskip
%%%%%%%%%%%%%%%%%%%%%%%%%%%%%%%%%%%%%%%%%%%%%%%%%%%%%%%%

\end{document}